# On the deprojection of axisymmetric bodies

Ortwin Gerhard[1] and James Binney[2]
[1] *Astronomisches Institut, Universität Basel, Venusstrasse 7, CH-4102 Binningen, Switzerland*
[2] *Theoretical Physics, Keble Road, Oxford OX1 3NP*

**ABSTRACT**

Axisymmetric density distributions are constructed which are invisible when viewed from a range of inclination angles $i$. By adding such distributions to a model galaxy, it can be made either disky or boxy without in any way affecting its projected image. As the inclination of a galaxy decreases from edge-on to face on, the range of 'invisible' densities, the uncertainty in the deprojection, and the sensitivity of the deprojection to noise all increase. The relation between these phenomena is clarified by an analysis of Palmer's deprojection algorithm.

These results imply that disk-to-bulge ratios are in principle ill-determined from photometry unless the disk is strong or the system is seen precisely edge-on. The uncertain role of third integrals in galaxies makes it unclear to what degree this indeterminacy can be resolved by kinematic studies.

**Key words:** Galaxies: fundamental parameters – galaxies: photometry – galaxies: kinematics and dynamics

## 1 INTRODUCTION

Since many galaxies are at least approximately axisymmetric and transparent, it is desirable to be able to estimate the three-dimensional luminosity density $\rho(r,\theta)$ of an axisymmetric galaxy from its projected surface brightness $I(x,y)$, given an assumed inclination angle $i$ between the galaxy's symmetry axis (the $z$-axis) and the line of sight to the observer. The Richardson–Lucy algorithm (Richardson 1972, Lucy 1974) has been successfully used to recover distributions $\rho(r,\theta)$ for a large number of galaxies with $i$ assumed different from 90° (e.g., Binney Davies & Illingworth 1990; van der Marel 1991; Dehnen 1995).

The uniqueness of the resulting model galaxies is unclear, however. On the one hand, Rybicki (1986) gave a simple argument based on the 'Fourier slice theorem' that, for $i \neq 90°$, any given surface brightness distribution, $I(x,y)$ must be the projection of an infinite number of luminosity densities $\rho(r,\theta)$. On the other hand, Palmer (1994) proved that the relationship between $I$ and $\rho$ is unique for $i \neq 0$ provided that the density $\rho(R,\theta)$ is 'band-limited': that is that its expansion in Legendre polynomials $P_l(\cos\theta)$,

$$\rho(r,\theta) = \sum_{l=0}^{L} \rho_l(r) P_l(\cos\theta), \qquad (1)$$

contains only a finite number of terms. Since any plausible luminosity density $\widetilde{\rho}$ can be approximated to sufficient accuracy by a band-limited density $\rho$, Palmer's result suggests that for practical purposes the relationship between $I$ and $\rho$ is one-to-one for all $i \neq 0°$.

If two space-densities can be found which, for fixed $i$, project to the same surface brightness distribution, then the difference between these densities projects to zero brightness. We refer to such an 'invisible' density distribution as a 'konus' density. Here we demonstrate by explicit construction of a family of konus densities that, for $i \neq 90°$, an infinite number of physically plausible luminosity densities *are* compatible with a given surface-brightness distribution. In particular, we show that one may choose between disky and boxy luminosity densities for the same photometric data.

We describe how the range of possible luminosity densities that are compatible with a given surface brightness distribution grows continuously as the inclination varies from edge-on to face-on, and how this change is reflected in the division of the space of possible luminosity densities $\rho(r,\theta)$ into its three parts: konus densities, densities that can be obtained by an extension of Palmer's inversion algorithm, and other visible densities.

In Section 2, we construct a family of konus densities and show that by adding one of these to a given deprojected brightness distribution we can turn a boxy distribution into a disky one, or vice versa. In Section 3 we show that Palmer's algorithm can provide a consistent deprojection of data that do not derive from a band-limited density distribution provided certain conditions on the data are satisfied. These conditions will be satisfied by most data when the assumed inclination angle is $i \simeq 90°$, and become more and more restrictive as $i \to 0$.

Section 4 discusses the effects of noise in the data, the division of the space of possible surface brightness distributions into ones that can be deprojected and ones that cannot, and the implications of the existence of konus densities for the detectability of low-luminosity disks in elliptical galaxies.

Section 5 sums up. In an appendix we give a group-theoretic proof of the result that resolves the apparent con-flict between the papers of Rybicki and Palmer; namely that the Fourier transform of a band-limited distribution is itself band-limited.

## 2 KONUS DENSITIES

We now show that there exist distinct three-dimensional light distributions that are both physically plausible – i.e., are smooth and sufficiently compact – and project to the same surface brightness distribution. This we do by explicitly calculating a family of axisymmetric luminosity distributions which are invisible when viewed at all inclination angles smaller than some critical value.

We start with a review of Rybicki's (1986) application of the Fourier slice theorem. By writing the axisymmetric density $\rho(\boldsymbol{x})$ in terms of its Fourier transform $A(\boldsymbol{k})$,

$$\rho(\boldsymbol{x}) = \frac{1}{(2\pi)^3}\int \mathrm{d}^3k\, A(\boldsymbol{k}) \exp(\mathrm{i}\boldsymbol{k}\cdot\boldsymbol{x}), \tag{2}$$

and projecting this along $z$, say,

$$\begin{aligned} I(x,y) &= \int \mathrm{d}z\, \rho(\boldsymbol{x}) \\ &= \frac{1}{(2\pi)^2}\int \mathrm{d}^3k\, A(k_x,k_y,0)\,\delta(k_z)\exp(\mathrm{i}\boldsymbol{k}\cdot\boldsymbol{x}), \end{aligned} \tag{3}$$

we see that the Fourier transform of the projected surface density $I(x,y)$ is the two-dimensional slice $A(k_x,k_y,0)$ of $A(\boldsymbol{k})$. This simple result is known as the Fourier slice theorem.

Consider now the projection of an axisymmetric density distribution. From the observed image we can obtain the two-dimensional slice of the density's Fourier transform with $\boldsymbol{k}$ perpendicular to the direction of projection. However, because of the assumed axial symmetry, all lines of sight which are related to the actual line of sight by a simple rotation around the symmetry axis must give identical images. Therefore by symmetry the Fourier transform is in fact known in all parts of Fourier space that are swept by rotating this two-dimensional slice around the object's symmetry axis.

If the density distribution is observed edge-on, this sweeping process covers *all* of Fourier space, and the three-dimensional density can be uniquely recovered from the projected image. In a face-on view, only a single Fourier plane is known even after the sweeping process, and, as is well-known, the deprojection is in this case highly non-unique. If, however, the direction of projection is inclined with respect to the symmetry axis by an angle $i \neq 90°$, then the rotation of the Fourier slice leaves a cone around the symmetry axis with half-angle $(90°-i)$ uncovered – the so-called 'cone of ignorance'. The surface of this cone opens to a plane (half-angle 90°) in the face-on case, and closes around the symmetry axis in an edge-on projection (half-angle 0°). Any density distribution derived from a Fourier transform which is non-zero only in the cone of ignorance will have zero projected brightness and thus be a konus distribution. Conversely, a density distribution whose Fourier transform is non-zero anywhere outside the cone of ignorance will have non-zero projected brightness. Hence konus densities are precisely those densities whose Fourier transforms are non-zero only in the cone of ignorance. Note that a konus density for inclination $i$ is also a konus density for any inclination $i' < i$, since the cone of ignorance for $i$ lies within the cones of ignorance for all smaller inclinations.

In the following we use Cartesian coordinates $(x,y,z)$ and cylindrical polar coordinates $(R,\phi,z)$ in the galaxy-intrinsic frame, with $z$ along the symmetry axis, and $(x',y',z')$ coordinates in the frame of the observer, such that projection is along $\boldsymbol{e}_{z'}$ and the $x=x'$-axis is the line of nodes. The inclination angle $i$ is measured between $\boldsymbol{e}_z$ and $\boldsymbol{e}_{z'}$, the $\boldsymbol{e}_{z'}$-axis having direction $(0,\sin i,\cos i)$ in the galaxy-intrinsic frame. The boundary of the cone of ignorance in $k$-space is given by

$$|k_z|/k_R \equiv |k_z|/\sqrt{k_x^2+k_y^2} = \cot(90°-i), \tag{4}$$

or

$$|k_z| = k_R \tan i. \tag{5}$$

The physical density corresponding to a Fourier density $A(\boldsymbol{k})$ in the cone of ignorance is then

$$\begin{aligned} \rho_{\rm K}(\boldsymbol{x}) = \frac{1}{(2\pi)^3}\int \mathrm{d}k_R\, k_R \int_{|k_z|\geq k_R\tan i} \mathrm{d}k_z\, \exp(\mathrm{i}k_z z) \\ \times \int_0^{2\pi} \mathrm{d}\phi_k\, \exp[\mathrm{i}Rk_R\cos(\phi_k-\phi)]\,A(\boldsymbol{k}), \end{aligned} \tag{6}$$

where $\boldsymbol{k} \equiv (k_R\cos\phi_k, k_R\sin\phi_k, k_z)$. By symmetry, $A(\boldsymbol{k})$ must be independent of $\phi_k$, so as not to generate a dependence of $\rho_{\rm K}(\boldsymbol{x})$ on $\phi$. Thus the $\phi_k$-integral simply evaluates to $2\pi J_0(k_R R)$, where $J_0$ is the usual Bessel function. We will, moreover, assume that $A(\boldsymbol{k})$ is a symmetric function with respect to $k_z=0$,

$$A(\boldsymbol{k}) = \tfrac{1}{2}\left[B(k_R,k_z) + B(k_R,-k_z)\right], \tag{7}$$

so that $\rho_{\rm K}(\boldsymbol{x})$ becomes the cosine-Bessel transform

$$\begin{aligned} \rho_{\rm K}(R,z) = \frac{1}{2\pi^2}\int_0^\infty \mathrm{d}k_R\, k_R J_0(k_R R) \\ \times \int_{k_R\tan i}^\infty \mathrm{d}k_z\, \cos(k_z z)\,A(k_R,k_z). \end{aligned} \tag{8}$$

### 2.1 A specific family of konus densities

From the discussion above it is clear that $A(\boldsymbol{k})$ must be zero on and outside the cone of ignorance. So that it lead to a physically reasonable space density we also require it to be smooth, and to have certain characteristic scales. This last requirement derives from the fact that a konus density must necessarily be positive in some parts of space and negative in others, and after adding a konus density to a visible one we want the resulting density to be everywhere non-negative. This requires that the konus density remain finite as $r \to 0$ and fall off sufficiently rapidly as $r \to \infty$. Consequently, the konus density must have certain characteristic scales, and these will be inversely related to corresponding scales in its Fourier density $A(\boldsymbol{k})$.

It is likely that a general investigation of Fourier densities that are confined to the cone of ignorance will be a difficult numerical problem. We have therefore sought an analytical example and, after some experimentation with Gradshteyn & Ryzhik (1980; hereafter GR), have concentrated on the following family of konus Fourier densities:

$$A(\boldsymbol{k}) = \begin{cases} \kappa^2 \exp(-\alpha\kappa)\, \exp(-\beta k_R) & \text{for } \kappa > 0, \\ 0 & \text{otherwise,} \end{cases} \tag{9}$$

where

$$\kappa \equiv |k_z| - k_R \tan i. \tag{10}$$

$A(\boldsymbol{k})$ decreases smoothly to zero on the boundary of the cone ($\kappa = 0$) and has its main contribution in a limited region of Fourier space around the $k_z$-axis, as determined by the two characteristic length scales $\alpha$ and $\beta$.

With this $A(k_R, k_z)$, the cosine-Bessel transform of (8) can be done simply. The $k_z$-integral gives

$$\begin{aligned} \int_{k_R \tan i}^{\infty} & \mathrm{d}k_z\, \cos(k_z z)\, \kappa^2 \exp(-\alpha\kappa) \\ &= \frac{\partial^2}{\partial\alpha^2} \frac{\alpha\cos(zk_R\tan i) - z\sin(zk_R\tan i)}{\alpha^2 + z^2} \\ &= g_1(z)\cos(zk_R\tan i) + g_2(z)\sin(zk_R\tan i), \end{aligned} \tag{11a}$$

where we have used formula 3.893.2 of GR and

$$\begin{aligned} g_1(z) &\equiv \frac{2\alpha^3 - 6\alpha z^2}{(\alpha^2+z^2)^3} \\ g_2(z) &\equiv \frac{2z^3 - 6\alpha^2 z}{(\alpha^2+z^2)^3}. \end{aligned} \tag{11b}$$

Note that for $z \gg \alpha$ the cosine and sine terms in this equation decrease like $\propto z^{-4}$ and $\propto z^{-3}$, respectively.

Before proceeding further with the wave density in eq. (9), it is instructive to replace the term $\exp(-\beta k_R)$ by $(2k_R)^{-1}\delta(k_R)$. Then the remaining integral can immediately be done, and we obtain the density of a plane-parallel sheet

$$\rho_s(z) = \frac{1}{2\pi^2} \frac{\alpha^3 - 3\alpha z^2}{(\alpha^2+z^2)^3}. \tag{12}$$

The surface density obtained on integrating along any direction with inclination $i' \neq 90°$ is

$$\begin{aligned} I_s(i') &= \int \mathrm{d}z' \rho_s(z) = \int \frac{\mathrm{d}z}{\cos i'} \rho_s(z) \\ &= \frac{1}{2\pi^2 \cos i'} \int_{-\infty}^{\infty} \mathrm{d}z\, \frac{\alpha^3 - 3\alpha z^2}{(\alpha^2+z^2)^3} = 0. \end{aligned} \tag{13}$$

Only when the sheet is seen edge-on ($i' = 90°$) is the massive wire along $k_z$ part of the slice in Fourier space about which information is available, and only in this case will the projected density be non-zero, with positive or negative values, depending on the vertical coordinate $y' = z$.

We now return to the exponential wave density (9). GR, formula 6.751.3, give the integral

$$\begin{aligned} C_0 &= \int_0^\infty \mathrm{d}k_R\, \exp(-\beta k_R)\, \cos(zk_R\tan i)\, J_0(k_R R) \\ &= \frac{\left(A + \sqrt{A^2+B}\right)^{1/2}}{\sqrt{2}\sqrt{A^2+B}}, \end{aligned} \tag{14}$$

where

$$\begin{aligned} A &\equiv R^2 - z^2\tan^2 i + \beta^2, \\ B &\equiv 4\beta^2 z^2 \tan^2 i. \end{aligned} \tag{15}$$

From this we may obtain the two $k_R$-integrals required in (8) as

$$\begin{aligned} C_1 &= \int_0^\infty \mathrm{d}k_R\, k_R \exp(-\beta k_R)\, \cos(zk_R\tan i)\, J_0(k_R R) \\ &= -\frac{\partial C_0}{\partial\beta} = \frac{\beta D_1 + 4\beta z^2 \tan^2 i\, D_2}{\sqrt{2}\, D_3} \end{aligned} \tag{16}$$

and, with $a \equiv z\tan i$,

$$\begin{aligned} S_1 &= \int_0^\infty \mathrm{d}k_R\, k_R \exp(-\beta k_R)\, \sin(zk_R\tan i)\, J_0(k_R R) \\ &= -\frac{\partial C_0}{\partial a} = \frac{z\tan i\,(-D_1 + 4\beta^2 D_2)}{\sqrt{2}\, D_3}, \end{aligned} \tag{17}$$

where

$$\begin{aligned} D_1 &\equiv A^2 + A\sqrt{A^2+B} - B, \\ D_2 &\equiv A + \tfrac{1}{2}\sqrt{A^2+B}, \\ D_3 &\equiv \left(A + \sqrt{A^2+B}\right)^{1/2} \left(A^2+B\right)^{3/2}. \end{aligned} \tag{18}$$

Thus, finally, the density corresponding to the Fourier density (9) may be written as

$$\rho_{\rm K}(R,z) = \frac{1}{2\pi^2} \left[g_1(z)\, C_1(R,z) + g_2(z)\, S_1(R,z)\right], \tag{19}$$

with the functions $g_i(z)$ defined by eqs (11b), and the remaining auxiliary quantities specified in eqs. (14) – (18).

This density is (i) everywhere non-singular, (ii) compact, and (iii) by construction, invisible at all inclinations smaller than $i$. To prove (i) we note from (15) that for $\beta \neq 0$ $A^2 \geq 0$ and $B \geq 0$, but both cannot be zero simultaneously. Thus the denominator $D_3$ of both $C_1$ and $S_1$ is positive definite. Also, for $\alpha \neq 0$ the denominator of both $g_1(z)$ and $g_2(z)$ is positive definite. To show (ii) we consider the asymptotic behaviour at large $R$ and $z$. For $R \to \infty$ at fixed $z$, we have $A \propto R^2$, $B \propto R^0$, $C_1 \propto A^{-3/2} \propto R^{-3}$, $S_1 \propto R^{-3}$ and, since $g_i$ are independent of $R$, also $\rho_{\rm K} \propto R^{-3}$. For $z \to \infty$ at fixed $R$, $A \propto z^2$, $B \propto z^2$, $C_1 \propto z^{-3}$, $S_1 \propto z^{-2}$, and $\rho_{\rm K} \propto g_2(z)\, S_1(R,z) \propto z^{-5}$. Finally, for the special direction given by $R^2 = z^2\tan^2 i$, we have $A = \beta^2$, $B = 4\beta^2 R^2$, and with $r^2 = R^2 + z^2$ $C_1 \propto r^{-1/2}$, $S_1 \propto r^{-1/2}$, $g_1 \propto r^{-4}$, $g_2 \propto r^{-3}$ and so $\rho_{\rm K} \propto r^{-7/2}$.

To confirm that this $\rho_{\rm K}(R,z)$ is invisible in projection for the inclination angle $i$ used in its construction, or indeed in any more face-on projection $i'$ with $i' \leq i$, we have integrated the density (19) numerically along lines of sight, using a Romberg midpoint rule algorithm and eq. (18) of Binney, Davies & Illingworth (1990). At all image points tried the projected density was found to be vanishingly small for $i' \leq i$, and then to rise relatively rapidly from zero for $i' \geq i$.

The konus densities (19) tend at large $R$ and small $z$ to constant-scale-height disks. To see this, we first note that for $R \to \infty$ and fixed $k_R$ the Bessel function in eqs. (16) and (17) is asymptotically

$$J_0(k_R R) \sim \left(\frac{2}{\pi k_R R}\right)^{1/2} \cos\left(k_R R - \frac{\pi}{4}\right) + \mathrm{O}\left(\frac{1}{k_R R}\right). \tag{20}$$

Thus as $R \to \infty$, only wavenumbers near $k_R = 0$ (very long wavelengths) contribute to the konus density (19), suggesting an asymptotic connection to the plane-parallel sheet solution (12). To make this more explicit, we rewrite the konus

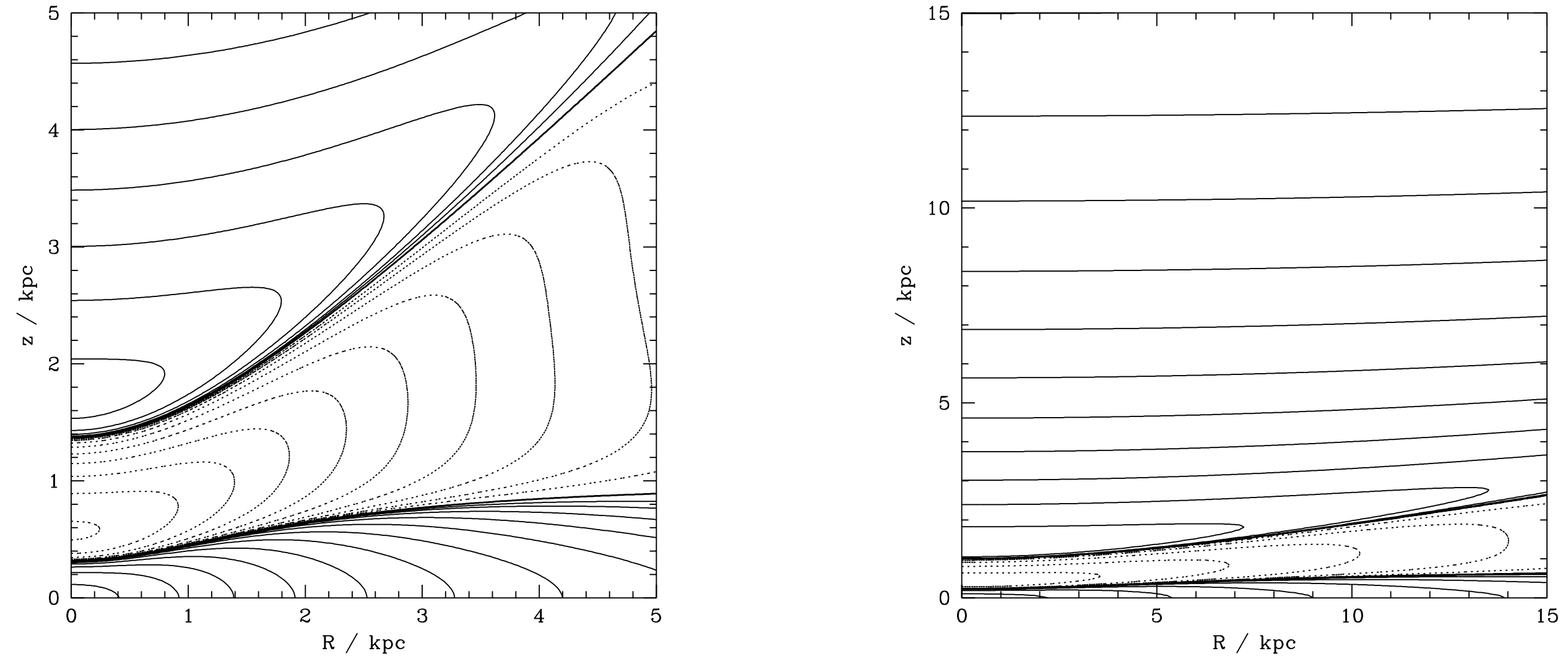

**Figure 1.** Contour plots in the meridional plane of two konus densities: $\alpha = \beta = 1$, $i = 45°$ (left panel); $\alpha = 0.75$, $\beta = 4$, $i = 80°$ (right panel). Contours are uniformly spaced in $\log(|\rho|)$ and for negative density are dotted.

density (19) as

$$\rho_{\rm K}(R,z) = \frac{1}{2\pi^2 R^2}\int_0^\infty {\rm d}x\, x J_0(x) \exp(-\beta x/R) \times \left[g_1(z)\cos\left(\frac{z}{R}x\tan i\right) + g_2(z)\sin\left(\frac{z}{R}x\tan i\right)\right]. \tag{21}$$

For $R \to \infty$, the konus density takes its largest values near the $R$-axis ($z \ll R$); at $z \sim R$ it is smaller by a factor of order $(\beta/R)^2$. Thus we may approximate the square bracket in the last equation by simply $g_1(z)$, after which the remaining integral gives (GR, 6.623.2)

$$\rho_{\rm K}(R,z) \simeq \frac{g_1(z)}{2\pi^2 R^2}\int_0^\infty {\rm d}x\, x J_0(x)\exp(-\beta x/R) = \frac{\beta}{\pi^2 R^3}\frac{\alpha^3 - 3\alpha z^2}{(\alpha^2+z^2)^3}, \tag{22}$$

which is the plane-parallel sheet solution (12) multiplied by a factor proportional to $R^{-3}$. Thus asymptotically for large $R$, the konus density (19) approaches a disk-like configuration that consists of a thin positive-density disk surrounded by a thicker layer of negative density.

## 2.2 Some illustrations

Fig. 1 shows contour plots in the meridional $(R, z)$ plane of two members of this family of konus densities. The first is constructed with parameters $\alpha = \beta = 1$ and $i = 45°$ in eq. (9), the second with $\alpha = 0.75$, $\beta = 4$, and $i = 80°$. Solid contours in Fig. 1 delineate regions in which the konus density is positive, dotted contours signify regions of negative density. The structure apparent in Fig. 1 is typical of all the konus densities we have generated from eq. (9) – a region of positive-density around the minor axis is followed by a region of negative density at intermediate latitudes, and then by another region of positive density around the $R$-axis. The approach as $R \to \infty$ of this last region to the plane-parallel sheet is apparent; at the largest radius plotted, the asymptotic density in the mid-plane, $\rho(R, 0) \simeq \beta/(\pi^2 R^3 \alpha^3)$ is accurate to better than a percent.

These density distributions project to zero surface brightness for all observers viewing them from inclinations $i' \leq 45°$ and $i' \leq 80°$, respectively – such lines of sight pass through regions of both positive and negative density. On the other hand, in an edge-on view ($i = 90°$) there are lines of sight which pass through only the region of positive density near the $R$-axis, or only the region of negative density above it, depending on height $z$. The structures in the right-hand panel of Fig. 1 appear to be flattened because for high inclination the cone of ignorance is narrow, allowing only for fairly long-wavelength components in the radial direction.

By adding or subtracting such density components to a 'visible' density distribution, such as those commonly inferred for elliptical galaxies, it is clearly possible to generate intrinsic disk-like or peanut-like components without altering the observed image. Ideally one would demonstrate this by choosing an underlying elliptical density distribution $\rho_E(R, z)$, such as a modified Hubble profile, say, and adding a variety of possible konus densities. Since here we have fixed the functional form of the konus density (19) a priori, we will instead vary the functional form of the elliptical density distribution. Fig. 2 shows three examples of such superpositions

$$\rho(R,z) = \rho_j(R,z) + f\,\rho_{\rm K}(R,z|\alpha,\beta,i). \tag{23}$$

In the top panel of Fig. 2 a boxy density distribution is obtained by subtracting a multiple of the 45°-konus density of Fig. 1 from the elliptical density distribution

$$\rho_3(m) \equiv \rho_0 m^{-1}(m+m_c)^{-2}. \tag{24}$$

Here $m$ is the usual spheroidal radius $m^2 \equiv R^2 + z^2/q^2$. This boxy distribution and the elliptical model (24) project to exactly the same elliptical isophotes for all inclination angles $i' \leq 45°$. The middle panel of Fig. 2 shows contours of a strongly disky intrinsic distribution, obtained by adding a multiple of the $\alpha=2$, $\beta=4$, 45°-konus density to the quasi-halo density

$$\rho_2(m) \equiv \rho_0(m^2+m_c^2)^{-1}. \qquad (25)$$

Again the isophotes of this disky model are precisely elliptical when viewed under inclination angles $i' \leq 45°$. Finally, the bottom panel of Fig. 2 shows the superposition of the quasi-halo model (25) with the 80°-konus density of Fig. 1; this shows that invisible disklike structures can be generated even for high inclination $i' \leq 80°$.

It is not entirely clear which of the features of the examples we have shown is generic and which determined by the peculiarities of the particular family of konus densities with which we have worked. Certainly, it would be straightforward to generate konus densities with different asymptotic profiles from the $\propto R^{-3}, z^{-5}$ characteristic of our family: densities with steeper asymptotic profiles could be generated by either differentiating eq. (19) w.r.t. $\beta$, or by replacing the exponential in eq. (9) by a Gaussian, say. Also, it is clear that any konus density must be made up of approximately conical shells of positive and negative density, and it seems unlikely that any has fewer than the three cones characteristic of our family. Konus densities with more cones certainly exist as one may demonstrate simply by adding konus densities of the family (19) for different values of $i$. For example, Fig. 3 shows the result of adding half the 45°-density of Fig. 1 to the 80°-density of Fig. 1. The combined konus density is invisible at $i \leq 45°$, and at $r > 4\,$kpc it has three conical regions of positive density and two regions of negative density. Adding such a density to an ellipsoidal model would not merely effect the transition from disky to boxy distributions.

The examples of Fig. 2 are significantly influenced by a combination of the $R^{-3}$ asymptotic profile and the fact that for $R \gg \beta$ the region of highest density around the $R$-axis tends to a constant-scale-height disk. Between them these limit the amplitude $f$ of the konus density which can be added to a spheroidal distribution without generating implausible isodensity surfaces.

Far from the origin, the structure of a konus density will be dominated by the behaviour of its Fourier transform near the origin of $k$-space. It happens that the particular wave-density (9) has a pronounced peak at $(k_R = 0, k_z = 2/\beta)$, with the result that waves that run parallel to the $z$-axis become dominant at large $r$. The tendency to a constant scale-height disk discussed above is a manifestation of this fact. In principle there is no reason why the Fourier transform of a konus density should not be zero along the $k_z$-axis. In this case the konus density would at large $R$ be dominated by waves with non-zero $k_R/k_z$, and would not tend to a constant scale-height disk. Some such Fourier transforms might lead to analytically tractable $k$-space integrals, or the integrals could be done numerically. Hence it is clear that the range of physically reasonable konus densities is certainly larger than that displayed by the family (19).

As the inclination increases towards edge-on, the angular scales of the konus densities decrease and the form of the 80°-konus density of Fig. 1 suggests that the available freedom in $\delta\rho/\rho$ is less at higher than at lower inclinations. However, because this statement is based on the detailed form of our konus densities, it is not well established. We shall return to this point with different arguments below.

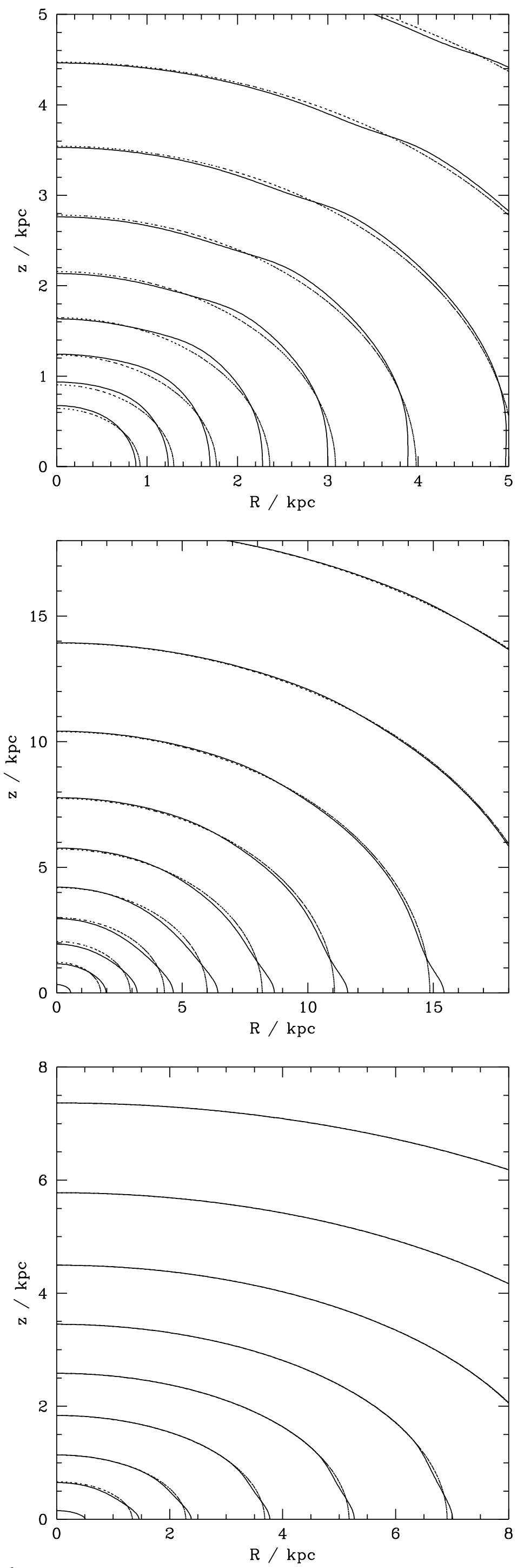

**Figure 2.** Disky and boxy systems with elliptical surface brightness contours. Each panel shows contours in the meridional plane for two deprojections of the same surface brightness data. The dotted elliptical contours correspond to one of the densities $\rho_j$ of eqs (24) or (25). The full contours show the sum of these densities and a konus density of plausible amplitude $f$. Top panel: $j=3$, $q=0.7$, $m_c=2$; $\alpha=\beta=1$, $i=45°$, $f<0$. Middle panel: $j=2$, $q=0.7$, $m_c=2$; $\alpha=2$, $\beta=4$, $i=45°$, $f>0$. Bottom panel: $j=2$, $q=0.5$, $m_c=3$; $\alpha=0.75$, $\beta=4$, $i=80°$, $f>0$.

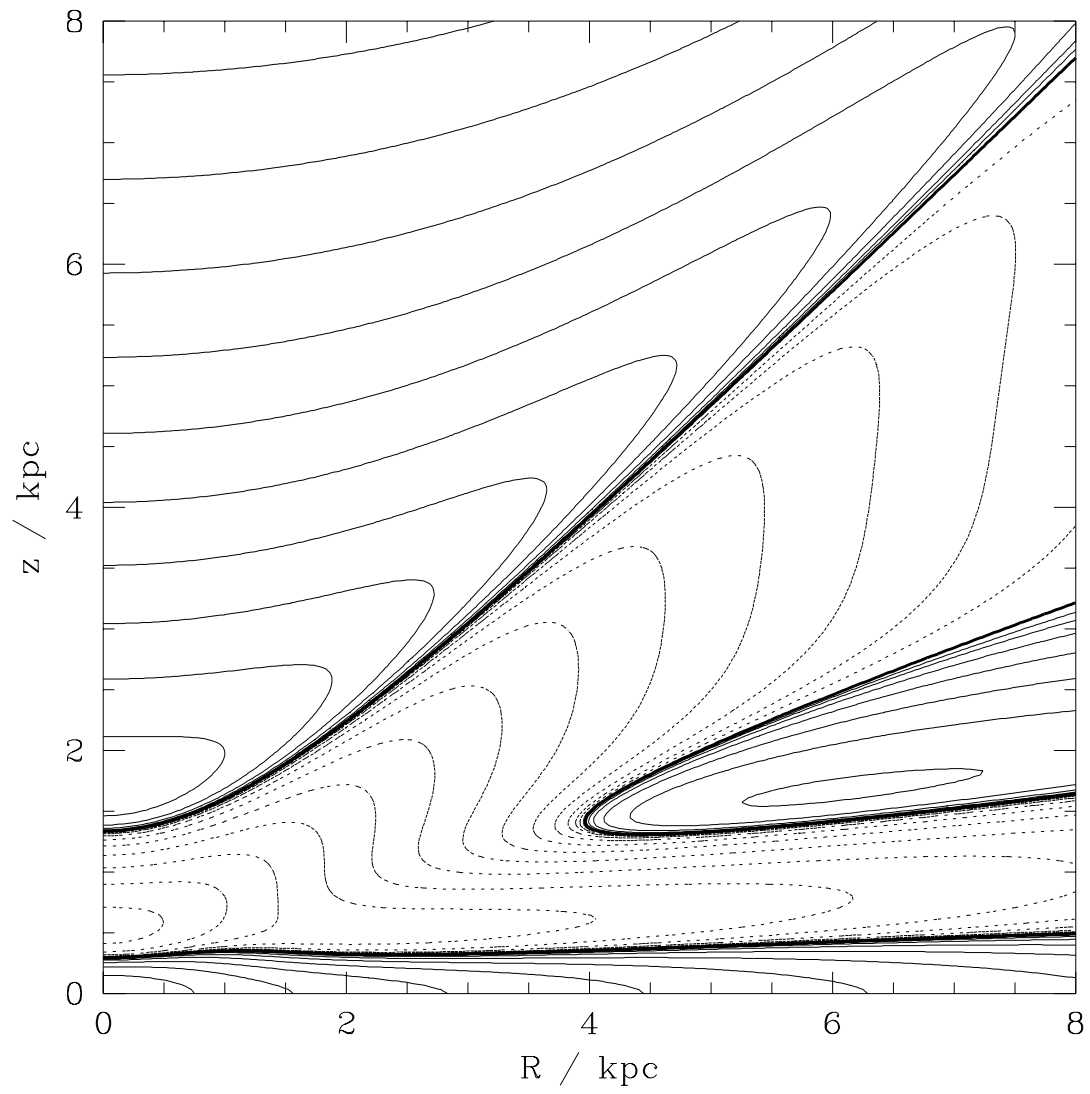

**Figure 3.** A linear combination of the two konus densities of Fig. 1: half the left panel has been added to the right panel. The resulting konus density has five conical regions and is invisible at inclinations $i \leq 45°$.

We have illustrated the effects of deprojection degeneracy by adding konus densities to intrinsically spheroidal distributions, whose isophotes are elliptical. But we could equally well have added konus densities to disky distributions that had pointed isophotes. Then we would have found that the amplitude and thickness of the disk implied by given data were highly ambiguous.

## 3 LEGENDRE EXPANSIONS

How do the results just presented relate to Palmer's (1994) proof that a band limited density distribution can be uniquely recovered from its projected surface brightness for all inclinations $i \neq 0$? To understand this we will first review Palmer's algorithm, and then return to the konus densities described above. We employ galaxy-intrinsic spherical polar coordinates $(r, \theta, \phi)$.

### 3.1 Palmer's algorithm

The Legendre polynomial expansion of a typical, smooth luminosity density $\widetilde{\rho}(r,\theta)$ will be the infinite sum

$$\widetilde{\rho}(r,\theta) = \sum_{l=0}^{\infty} \widetilde{\rho}_l(r) P_l(\cos\theta). \tag{26}$$

For simplicity we assume that $\widetilde{\rho}$ is symmetric about the equatorial plane, with the result that only even $l$ need be included in the sum. Let $(s,\varphi)$ be suitably oriented polar coordinates for the plane of the sky. Then the projection $\widetilde{I}$ of $\widetilde{\rho}$ may be written

$$\widetilde{I}(s,\varphi) = \tfrac{1}{2} I_0 + \sum_{n=2,4\ldots}^{\infty} I_n(s) \cos n\varphi. \tag{27}$$

If $\widetilde{\rho}$ were band-limited, the sum over $n$ in (27) would run only up to $n = L$, the largest value of $l$ involved in the Legendre polynomial expansion of $\widetilde{\rho}$. Thus the Fourier decomposition of the projection of a band-limited density distribution is band-limited in the conventional sense.

We consider the band-limited approximations $\rho^{(L)}$ to $\widetilde{\rho}$ that are obtained by setting to zero all the $I_n$ with $n > L$. Let the Legendre polynomial expansion of $\rho^{(L)}$ be

$$\rho^{(L)} = \sum_{l=0}^{L} \rho_l^{(L)}(r) P_l(\cos\theta). \tag{28}$$

By expressing the $(r,\theta,\phi)$ coordinates natural to the galaxy in terms of coordinates $(s,\varphi,z')$ natural to the observer, and integrating along $z'$, Palmer shows that

$$I_n(s) = 4 \sum_{\substack{l=n \\ (l-n)\ \mathrm{even}}}^{L} p_l^n(\cos i) \int_s^{\infty} \rho_l^{(L)}(r)\, p_l^n(\mu) \frac{r\,\mathrm{d}r}{\sqrt{r^2 - s^2}}, \tag{29}$$

where $p_l^n$ is related to the conventional associated Legendre function $P_l^n$ by

$$p_l^n \equiv \sqrt{\frac{(l-|n|)!}{(l+|n|)!}} P_l^n \quad \text{and} \quad \begin{aligned} \mu &\equiv \cos\big(\arcsin(s/r)\big) \\ &= \sqrt{1 - s^2/r^2}. \end{aligned} \tag{30}$$

[With this definition, $\int_{-1}^{1} \mathrm{d}x\,(p_l^n)^2 = 2/(2l+1)$.] Palmer shows that when $i \neq 0$, equations (29) can be converted into expressions for the $\rho_l^{(L)}$ in terms of the $I_n$, which may be determined observationally. Specifically,

$$\rho_l^{(L)}(r) = -\frac{2^l l! [(2l)!]^{-1/2}}{2\pi r p_l^l(\cos i)} \int_1^{\infty} \frac{\mathrm{d}}{\mathrm{d}\sigma}\left[\frac{\hat{I}_l^{(L)}(\sigma r)}{\sigma^l}\right] \frac{\mathrm{d}\sigma}{\sqrt{\sigma^2 - 1}}, \tag{31}$$

where

$$\begin{aligned} \hat{I}_l^{(L)}(s) \equiv I_l(s) &- 4 \sum_{l'=l+2}^{L} p_{l'}^l(\cos i) \\ &\times \int_s^{\infty} \rho_{l'}^{(L)}(r) p_{l'}^l(\mu) \frac{r\,\mathrm{d}r}{\sqrt{r^2 - s^2}}. \end{aligned} \tag{32}$$

Palmer's proof that band-limited densities can be uniquely inverted follows from the fact that for $i \neq 0$ this system of equations can be straightforwardly solved: one first finds $\rho_L^{(L)}$, which depends only on $I_L$. Then one solves for $\rho_{L-2}^{(L)}$, which depends on $I_{L-2}$ and $\rho_L^{(L)}$, and so on down the series of the $\rho_l^{(L)}$.

### 3.2 Extension to non-band-limited data

We now investigate under what circumstances Palmer's algorithm can be used to deproject a surface-brightness distribution $\widetilde{I}(s,\phi)$ that is not band-limited. Consider the difference between the band-limited approximations $\rho^{(L+2)}$ and $\rho^{(L)}$ to $\widetilde{\rho}$ that are obtained with Palmer's method by truncating the Fourier series (27) for $\widetilde{I}$ at $n = L+2$ and $n = L$, respectively. We have that

$$\begin{aligned} \delta\rho_l &\equiv \rho_l^{(L+2)} - \rho_l^{(L)} \\ &= -\frac{2^l l! [(2l)!]^{-1/2}}{2\pi r p_l^l(\cos i)} \int_1^{\infty} \frac{\mathrm{d}}{\mathrm{d}\sigma}\left[\frac{\delta\hat{I}_l(\sigma r)}{\sigma^l}\right] \frac{\mathrm{d}\sigma}{\sqrt{\sigma^2 - 1}}, \end{aligned} \tag{33}$$

where for $l = L+2$

$$\delta\hat{I}_{L+2} = I_{L+2}, \tag{34}$$

and for $l \le L$

$$\delta\hat I_l(s) = -4\sum_{l'=l+2}^{L+2} p^l_{l'}(\cos i)\int_s^\infty \delta\rho_{l'}(r)\,p^l_{l'}(\mu)\frac{r\,\mathrm{d}r}{\sqrt{r^2-s^2}}. \qquad (35)$$

Consider now the way in which $\delta\rho_l$ depends upon the signal $I_{L+2}$ under the assumption that the $I_n$ are slowly varying functions of $s$. For large $l$, the integral in equation (33) may be estimated as follows. We have

$$\frac{\mathrm{d}}{\mathrm{d}\sigma}\left[\frac{\delta\hat I_l(\sigma r)}{\sigma^l}\right] \simeq -\frac{l\,\delta\hat I_l(\sigma r)}{\sigma^{l+1}}, \qquad (36)$$

and the dominant contribution to the integral comes from $\sigma \simeq 1$, so

$$\begin{aligned}\int_1^\infty \frac{\mathrm{d}}{\mathrm{d}\sigma}\left[\frac{\delta\hat I_l(\sigma r)}{\sigma^l}\right]\frac{\mathrm{d}\sigma}{\sqrt{\sigma^2-1}} &\simeq -l\beta_l\delta\hat I_l(r)\int_1^\infty \frac{\mathrm{d}\sigma}{\sigma^{l+1}\sqrt{\sigma^2-1}} \\ &\simeq -l\,\frac{(l-1)!!}{l!!}\,\frac{\pi}{2}\beta_l\delta\hat I_l(r),\end{aligned} \qquad (37)$$

where $\beta_l$ is a number of order unity. Thus from equation (33)

$$|\delta\rho_l| \simeq l\,\frac{(l-1)!!}{l!!}\sqrt{\frac{(2l)!!}{(2l-1)!!}}\,\frac{\beta_l\delta\hat I_l(r)}{4rp^l_l(\cos i)}. \qquad (38)$$

Note that since $l = l!!(l-1)!!/[(l-1)!!(l-2)!!)]$, we have for large $l$ that $l!!/(l-1)!! \simeq \sqrt{l}$. Hence for large $l$,

$$|\delta\rho_l(r)| \sim l^{3/4}\,\frac{\delta\hat I_l(r)}{rp^l_l(\cos i)}. \qquad (39)$$

Similarly, the integral in equation (35) can be estimated as

$$\begin{aligned}\int_s^\infty &\delta\rho_{l'}(r)\,p^l_{l'}(\mu)\frac{r\,\mathrm{d}r}{\sqrt{r^2-s^2}} \\ &\simeq \alpha_{l'}\delta\rho_{l'}(s)\int_s^\infty p^l_{l'}\left(\sqrt{1-s^2/r^2}\right)\frac{r\,\mathrm{d}r}{\sqrt{r^2-s^2}} \\ &= \alpha_{l'}\,s\,\delta\rho_{l'}(s)\,\frac{(-1)^l\pi}{2^{l+1}l!}\sqrt{\frac{(l'+l)!}{(l'-l)!}}\,\frac{(l-3)!!}{(l-2)!!} \\ &\quad\times {}_3F_2\big(\tfrac{l'+l+1}{2}, -\tfrac{l'-l}{2}, \tfrac{l-1}{2}, l+1, \tfrac{l}{2}; 1\big).\end{aligned} \qquad (40)$$

where we have used equation 7.132.6 of GR. Here ${}_3F_2$ is the generalized hypergeometric series, which in this case terminates after $\frac{1}{2}(l'-l)+1$ terms. In general we find that a reasonable approximation to the resulting scaling of $\delta\hat I_l$ is

$$\begin{aligned}|\delta\hat I_l(s)| &\sim \sum_{l'=l+2}^{L+2}\frac{p^l_{l'}(\cos i)\,s\,\delta\rho_{l'}(s)}{l'\,(l/l')^2} \\ &\sim \sum_{l'=l+2}^{L+2}\frac{1}{l'^{1/4}}\left(\frac{l'}{l}\right)^2\frac{p^l_{l'}(\cos i)}{p^{l'}_{l'}(\cos i)}\,\delta\hat I_{l'}(s).\end{aligned} \qquad (41)$$

This equation allows us to estimate the effect on the recovered density distribution of adding one more term $I_{L+2}$ to the expansion of the surface brightness: every coefficient $\rho_l$ depends through (31) on the corresponding 'effective data' coefficient $\hat I_l$, and all of these are modified by the addition of $I_{L+2}$ in a way that we now estimate from (41). By equation (34), $\delta\hat I_{L+2} = I_{L+2}$, so from (41) we have

$$\delta\hat I_L \sim \frac{1}{(L+2)^{1/4}}\left(\frac{L+2}{L}\right)^2\frac{p^L_{L+2}(\cos i)}{p^{L+2}_{L+2}(\cos i)}\,I_{L+2}. \qquad (42\text{a})$$

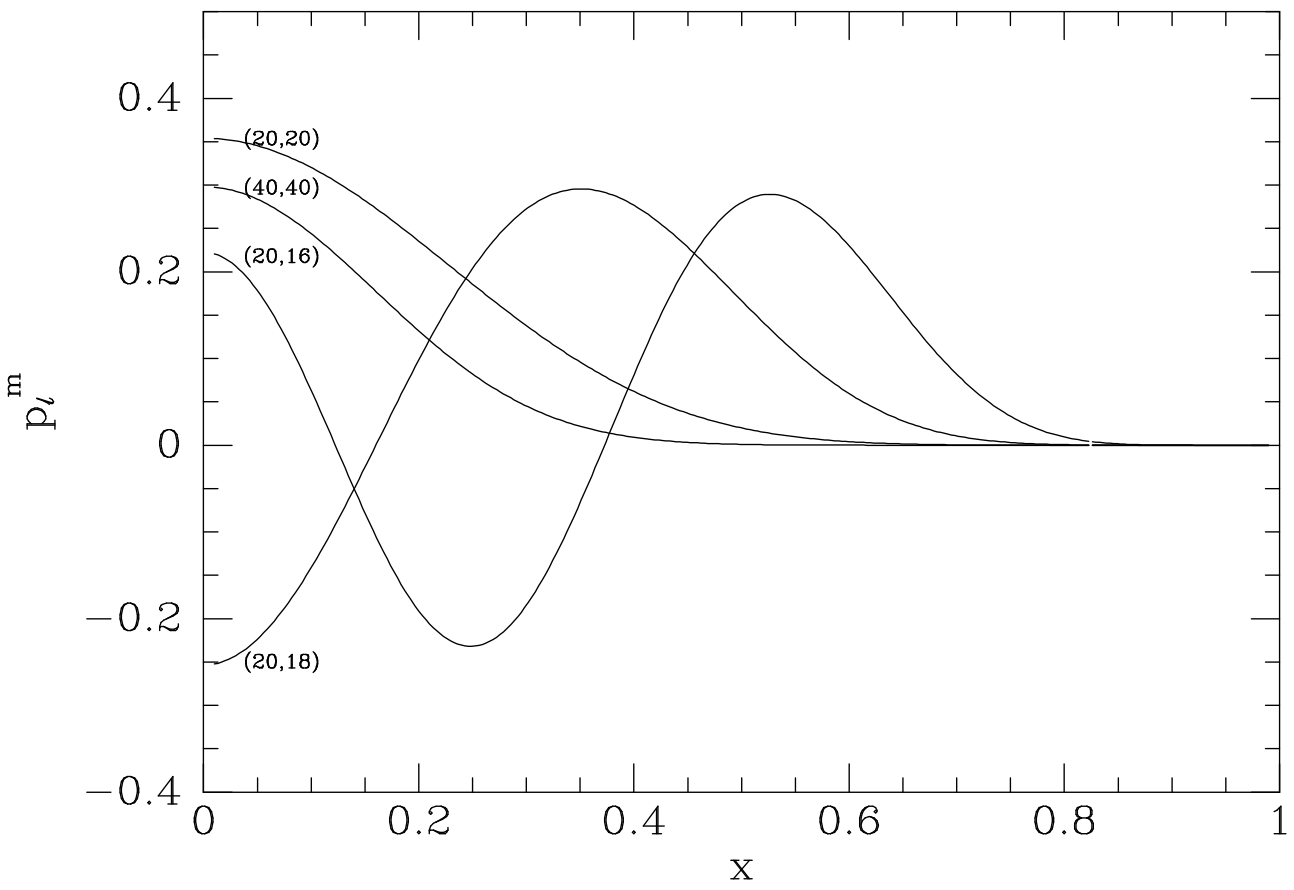

**Figure 4.** Typical Legendre functions normalized according to equation (30).

Using this to eliminate $\delta\hat I_L$ from the corresponding expression for $\delta\hat I_{L-2}$, we find

$$\begin{aligned}\delta\hat I_{L-2} \sim \Bigg[&\frac{1}{(L+2)^{1/4}L^{1/4}}\,\frac{p^L_{L+2}(\cos i)}{p^{L+2}_{L+2}(\cos i)}\,\frac{p^{L-2}_L(\cos i)}{p^L_L(\cos i)} \\ &+\frac{1}{(L+2)^{1/4}}\,\frac{p^{L-2}_{L+2}(\cos i)}{p^{L+2}_{L+2}(\cos i)}\Bigg]\left(\frac{L+2}{L-2}\right)^2 I_{L+2}.\end{aligned} \qquad (42\text{b})$$

The pattern of coefficients obtained by continuing this process down to arbitrary $l$ is now apparent. The dependence of $\delta\hat I_l$ on $I_{L+2}$ is governed by the ratios of the type $p^{l'-2k}_{l'}/p^{l'}_{l'}$. Fig. 4 shows three typical Legendre functions. From this one sees that the functions $p^{l'}_{l'} \propto \sin^{l'} i$ that occur in the denominators above fall monotonically as the inclination changes from edge-on to face-on, and become very small by a value of $x \equiv \cos i$ which diminishes as $l'$ increases. The functions $p^{l'-2k}_{l'}$ that appear in the numerators above oscillate for small $\cos i$, and tend to zero $\sim \sin^{(l'-2k)} i$ as $\cos i \to 1$. Consequently, the ratios of Legendre functions in (42) will be of order unity near $i = 90°$ but be large for near face-on inclinations. In fact, in the limit $i \to 0$ the products of Legendre functions in each term in the series of (42) will all be comparable because they all scale as $\sin^{-(L+2-l)} i$. For moderate inclinations and values of $L$, the ratio $\delta\hat I_l/I_{L+2}$ will be dominated by these products. Hence the addition of a small coefficient $I_{L+2}$ can profoundly modify the effective data $\hat I_l$ at all orders, and hence significantly change the recovered density distribution.

As we have defined it, $\delta\hat I_l$ is a function of the order $L$ of the last included term. If deprojection is to make sense, the difference between the deprojections that one obtains on truncating the data at either order $L$ or order $L+2$ must tend to zero as $L \to \infty$. Consequently, we require $\lim_{L\to\infty}\delta\hat I_l = 0$ for all $l \ll L$. For $i = 90°$, the ratios of Legendre functions in equation (42) will evaluate to less than unity, so $\lim_{L\to\infty}\delta\hat I_l$ is zero provided the Fourier coefficients $I_L$ fall off at least as fast as $L^{-2}$, as they will because the surface brightness is a continuous function of position angle on the sky. However, for small $i$, the limit will vanish only if the coefficients $I_L$ fall off extremely rapidly since they must overwhelm the very rapid growth with $L$ of the products of Legendre functions.

### 3.3 An example

It is instructive to see how these ideas work out in practice. Consider deprojecting a system whose noise-free surface brightness follows the modified Hubble profile

$$I(x,y)=\frac{I_0}{\mu_2^2}\quad\text{where}\quad \mu_2^2\equiv 1+\frac{x^2+y^2/q_2^2}{r_0^2}. \tag{43}$$

The luminosity density of this system is

$$\rho(r,\theta)=\frac{\rho_0}{\mu_3^3}\quad\text{where}\quad \mu_3^2\equiv 1+\frac{r^2}{r_0^2}\left(\sin^2\theta+\cos^2\theta/q_3^2\right) \tag{44}$$

with

$$\begin{aligned}\rho_0&=\frac{q_2 I_0}{2q_3 r_0},\\ q_3^2&=q_2^2\csc^2 i-\cot^2 i.\end{aligned} \tag{45}$$

By straightforward contour integration one may show that for this system

$$I_L(s)=\frac{2I_0}{\frac{1}{2}(q_2^{-2}+1)s^2+r_0^2}\frac{\left(\frac{\sqrt{1-\alpha^2}-1}{\alpha}\right)^{L/2}}{\sqrt{1-\alpha^2}}, \tag{46}$$

where $L$ is even and

$$\alpha(s)\equiv-\frac{q_2^{-2}-1}{(q_2^{-2}+1)+2(r_0/s)^2}. \tag{47}$$

Since

$$\begin{aligned}\left(\frac{\sqrt{1-\alpha^2}-1}{\alpha}\right)&\simeq\frac{1-q_2}{1+q_2}\quad(r\gg r_0)\\ &=\frac{\sin^2 i(1-q_3^2)}{\left(1+\sqrt{q_3^2\sin^2 i+\cos^2 i}\right)^2}.\end{aligned} \tag{48}$$

we have that $I_L$ tends to zero with $L$ as

$$I_L\sim\sin^L i(1-q_3^2)^{L/2}. \tag{49}$$

When this is inserted into equations (42) for $\delta\hat{I}_l$, the factor $\sin^{L+2} i$ in $I_{L+2}$ overwhelms the factors $\sin^{-(L+2-l)} i$ implicit in the products of Legendre functions, with the result that $\delta\hat{I}_l\sim\sin^l i$ just as $I_l\sim\sin^l i$. Thus Palmer's algorithm when used to deproject a noise-free modified Hubble profile might be expected always to converge to the same density distribution, independent of the inclination at which the system is viewed.

Figure 5 shows a numerical verification of this proposition. The open symbols show for a typical radius values of $\log_{10}|\rho_l|$ that were recovered by deprojecting a modified-Hubble system of axis ratio $q_3=0.6$ and inclination $i=10°$. The open symbols give the values obtained for $L=4,6$ and 8, while the full symbols give the values that one obtains by directly expressing $\rho(r,\theta)$ as a Legendre series. It can be seen that as $L$ is increased, good values are obtained for more and more of the $\rho_l$. A virtually indistinguishable figure could be shown for any other inclination angle $i>0$. Thus in this case, Palmer's algorithm does seem to recover the correct three-dimensional density from its projected surface brightness distribution.

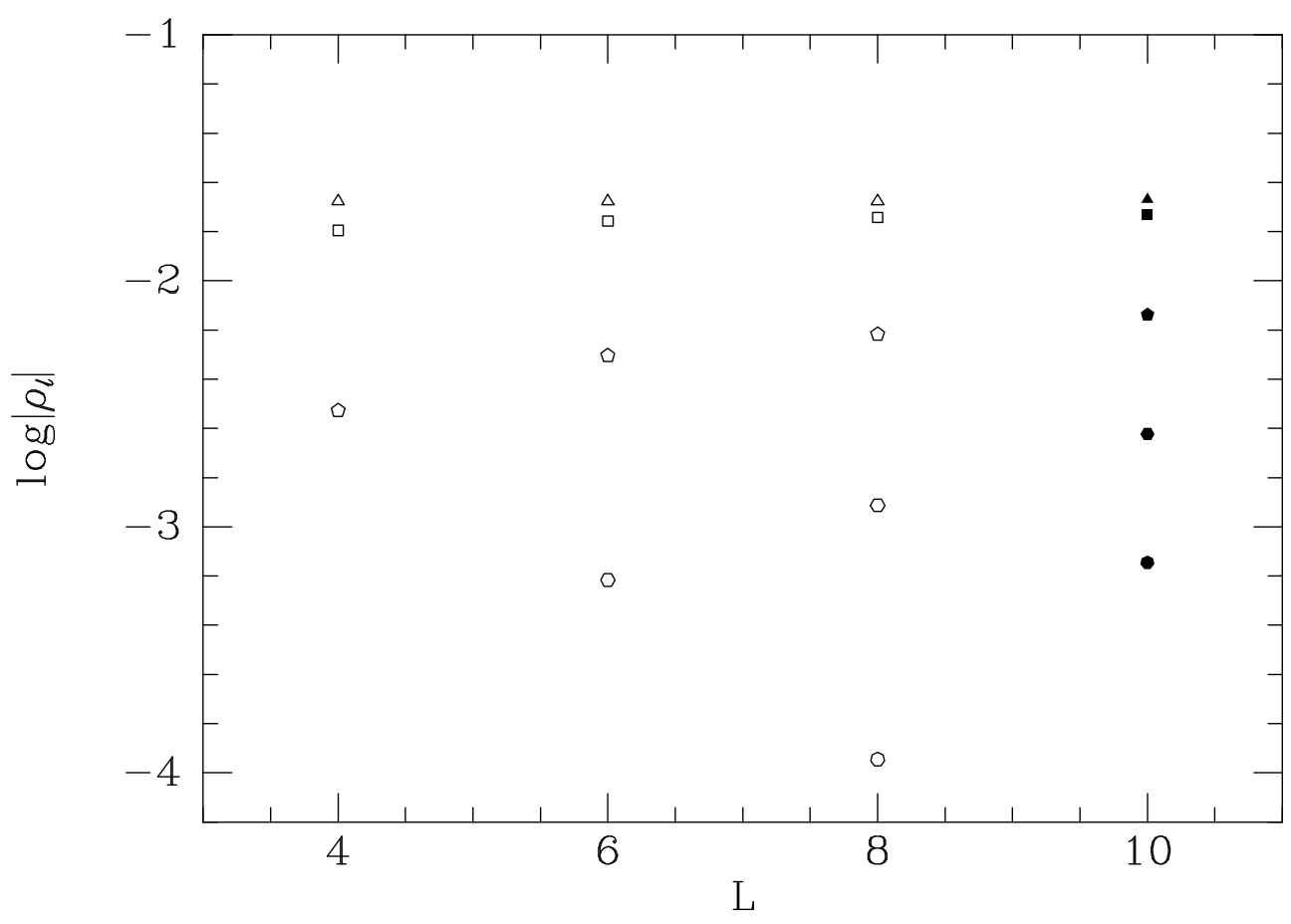

**Figure 5.** Legendre coefficients for a galaxy that has the modified Hubble profile. An E4 galaxy is viewed at $i=10°$. The full symbols show the values of $\log_{10}|\rho_l|$ for this density distribution at a typical radius. The open symbols show the values of the same quantities that are recovered from the projected density for $L=4,6,8$. Triangles give values of $\rho_0$, squares of $\rho_2$, pentangles of $\rho_4$, and so forth.

### 3.4 Konus densities and Palmer's algorithm

We first demonstrate that the Legendre expansion of a konus density must contain an infinite number of terms; i.e., it cannot be band-limited. Indeed, if $\rho_l(r)=0$ for $l>L$, then by (29) it follows from the fact that $I_L(s)=0$ for all $s$, that $\rho_L(r)=0$ for all $r$. Repeating this argument with $n=L-1,\ldots$ in (29) we see that a band-limited density distribution with zero surface brightness is identically zero. Fig. 6 illustrates this result by showing for one of the konus densities of Fig. 1 the coefficients $\rho_l(r)$ with $l\le 250$ at six values of $r$.

By adding a konus density to a band-limited distribution, we see that a band-limited surface brightness $I(s,\phi)$ does *not* imply that the underlying density $\rho(r,\theta)$ is band-limited, while eq. (29) clearly shows that the reverse is true.

Since, by construction, a konus density projects to zero surface brightness, one might think it impossible to recover such a density by Palmer's deprojection algorithm. Consider, however, the result of truncating the Legendre expansion of a konus density at order $L$. This band-limited density distribution projects to a band-limited surface brightness distribution $I^{(L)}(s,\phi)$. For any finite $L$ we may in principle deproject this by Palmer's algorithm. The band-limited result of this deprojection must coincide with our original truncated konus density, for otherwise the difference between these two densities would be a band-limited konus density, which we have just shown to be impossible. Repeating this experiment for larger and larger values of $L$, we obtain more and more accurate approximations to the original konus density, and in the limit $L\to\infty$ we obtain the konus density itself. Notice that this implies

$$\lim_{L\to\infty}\mathrm{Pa}\big(I^{(L)}\big)\neq\mathrm{Pa}\big(\lim_{L\to\infty}I^{(L)}\big), \tag{50}$$

where Pa stands for Palmer's operator. Indeed, as $L$ increases, $I^{(L)}$ becomes smaller and smaller and vanishes in the limit $L\to\infty$. Hence Palmer's algorithm extracts better and better approximations to the konus density's $\rho_l$ from smaller and smaller values of $I_l$. Equations (42) give some insight into how this is achieved, although they describe a limiting process that is different from the one under consideration here: now with each increment in $L$, *every* $I_l$ varies, rather than just $I_{L+2}$. In other words, we obtain a konus density from a series of Fourier series rather than from a single Fourier series.

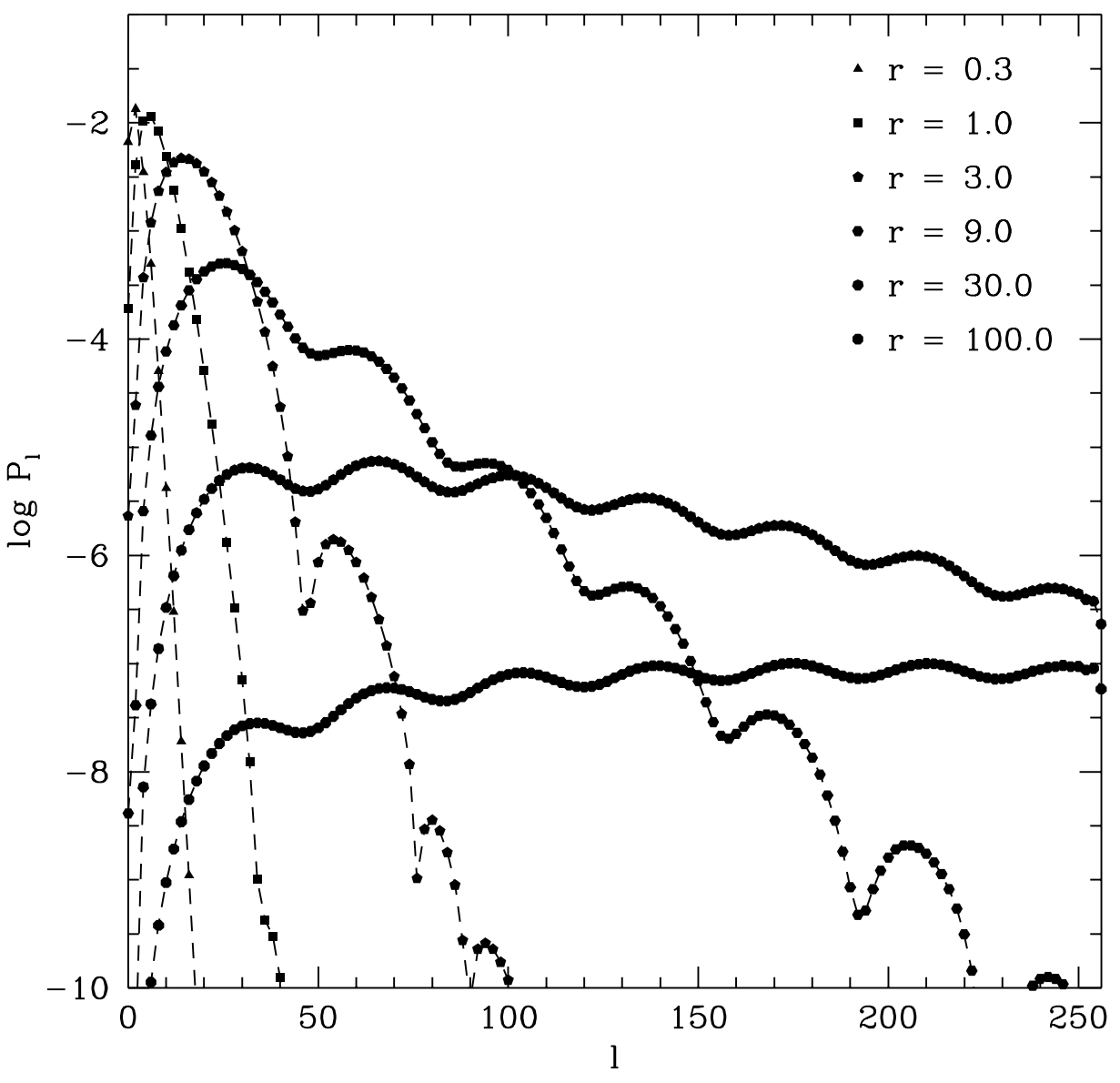

**Figure 6.** The coefficients $\rho_l(r)$ with $l \leq 250$ for the 80°-konus density of Fig. 1. The larger the value of $r$, the more slowly $\rho_l$ declines with $r$. This reflects the fact that at large $R$ these distributions approximate disks of constant scale-height. The 45°-konus density yields a similar figure.

## 4 DISCUSSION

### 4.1 Noise

We have seen that equations (42) impose constraints on the behaviour of the Fourier coefficients $I_l$ at large $l$ that must be satisfied if a meaningful deprojection is to be obtained by Palmer's algorithm. When the surface brightness distribution $I(s,\phi)$ is contaminated by noise, these conditions must be violated at sufficiently large $l$, since noise will tend to make $|I_l|$ approximately independent of $l$ for large $l$. Moreover, as the assumed inclination $i$ decreases, the conditions imposed by equations (42) become ever more severe, with the result that the value of $l$ at which a given body of noisy data will first violate these conditions diminishes with $i$.

In practice galaxy images are deprojected using the Richardson–Lucy (R–L) algorithm rather than Palmer's. Experience shows that for a given image there is a smallest value of $i$ for which the R–L algorithm yields a plausible density distribution. When using the R–L algorithm, both images and density distributions are usually fitted by low-order functions of the angles $\phi$ and $\theta$. This smoothing process will strongly attenuate the high-frequency power in both data and model, but it will not amount to strict truncation of the implicit Fourier and Legendre expansions. Consequently, the R–L algorithm will not be working with band-limited data or models. Equations (42) indicate that at sufficiently small values of $i$, even the residual small high-$l$ terms $I_l$ in the data can make important contributions to the low-$l$ coefficients $\rho_l$. Since the high-order $I_l$ will be noise dominated, it follows that the inversion will be entirely noise-dominated for sufficiently small assumed inclinations. Clearly, the deleterious effects of noise can only be *increased* by increasing the angular resolution at which the data are represented.

We have seen that the low-amplitude brightness distributions of truncated konus densities deproject to densities of amplitude unity. Any component of residual noise in smoothed data that is equal to the surface brightness of a truncated konus density can be matched *only* by projecting a truncated konus density. Hence no matter how such data are fitted, a model that fits the data accurately must include a truncated konus density of significant amplitude. Moreover, the less the data have been smoothed, and therefore the higher the value of $L$ at which the underlying series are truncated, the smaller is the amplitude of noise in the data that is required to produce a konus density of unit amplitude.

It is interesting to study a practical example of a truncated konus density being introduced into a model by noise in the data. Fig. 7 shows a deprojection of the surface brightness distribution of NGC 2300 for an assumed inclination angle of $i = 50°$. This galaxy is approximately E2, and has slightly boxy ($a_4 \lesssim 1\%$) isophotes. The deprojection was done with the R–L algorithm of Binney, Davies & Illingworth (1990) as implemented by W. Dehnen. After 10 iterations the rms deviation in surface brightness for six radial rays was 0.006 mag. Despite this very accurate representation of the nearly elliptical isophotes, the contours of deprojected density in the meridional plane are strongly non-elliptical and their shapes vary with radius. The shapes of these contours suggest that a truncated konus density has been added to a smooth elliptical model.

Ambiguities in the result of the deprojection can be avoided by introducing additional constraints, such as that the deprojected density distribution should be as smooth as possible, or as elliptical as possible, etc. However, the realization that many elliptical galaxies display fine structure such as weak disks casts doubts on the wisdom of such prescriptions. The successes of the R–L algorithm referred to in the introduction may in part be due to its effectively smoothing the angular density distribution during the deprojection, thus removing the higher $P_l(\cos\theta)$.

### 4.2 What can and cannot be seen

We have made it plausible that Palmer's algorithm can be applied to some non-band-limited surface brightness distributions in addition to band-limited ones. For any assumed inclination $i$, let $2\mathrm{D_P}(i)$ denote the set of surface-brightness distributions $I$ which yield a well-defined deprojected density distribution in the limit $L \to \infty$ as Palmer's algorithm is successively applied to the Fourier decomposition truncated at each order $L$. Then $2\mathrm{D_P}(i)$ comprises nearly all surface-brightness distributions for $i = 90°$ and shrinks continuously with $i$ until at $i = 0$ it comprises only circularly-symmetric brightness distributions. Thus, whereas in Palmer's original discussion the case $i = 0$, in which Palmer's algorithm cannot be applied even to band-limited distributions, appeared anomalous, we now see that it is merely the endpoint of a continuous evolution.

For any given value of $i$, there is a set $3\mathrm{D_P}(i)$ of density distributions that can be obtained by applying Palmer's algorithm to all the members of $2\mathrm{D_P}(i)$. One expects that the set $3\mathrm{D_P}(i)$ of functions of two variables will have essentially the same size as the set $2\mathrm{D_P}(i)$ of similar functions of which it is the image. No konus density lies in $3\mathrm{D_P}(i)$ because to recover a konus density by Palmer's algorithm one has to change all the Fourier coefficients $I_n$ at each stage of the limiting process, rather than adding one more term to the Fourier series. Hence the set of konus densities, $3\mathrm{D_K}(i)$, is disjoint from $3\mathrm{D_P}(i)$. On adding a member of $3\mathrm{D_K}(i)$ to a member of $3\mathrm{D_P}(i)$ we obtain a density distribution that lies in neither set. Hence the space of all possible density distributions falls into three parts as is schematically illustrated by Fig. 8. That figure also illustrates how $3\mathrm{D_K}(i)$ increases from the empty set to a large part of the space as $i$ diminishes from 90° to zero.

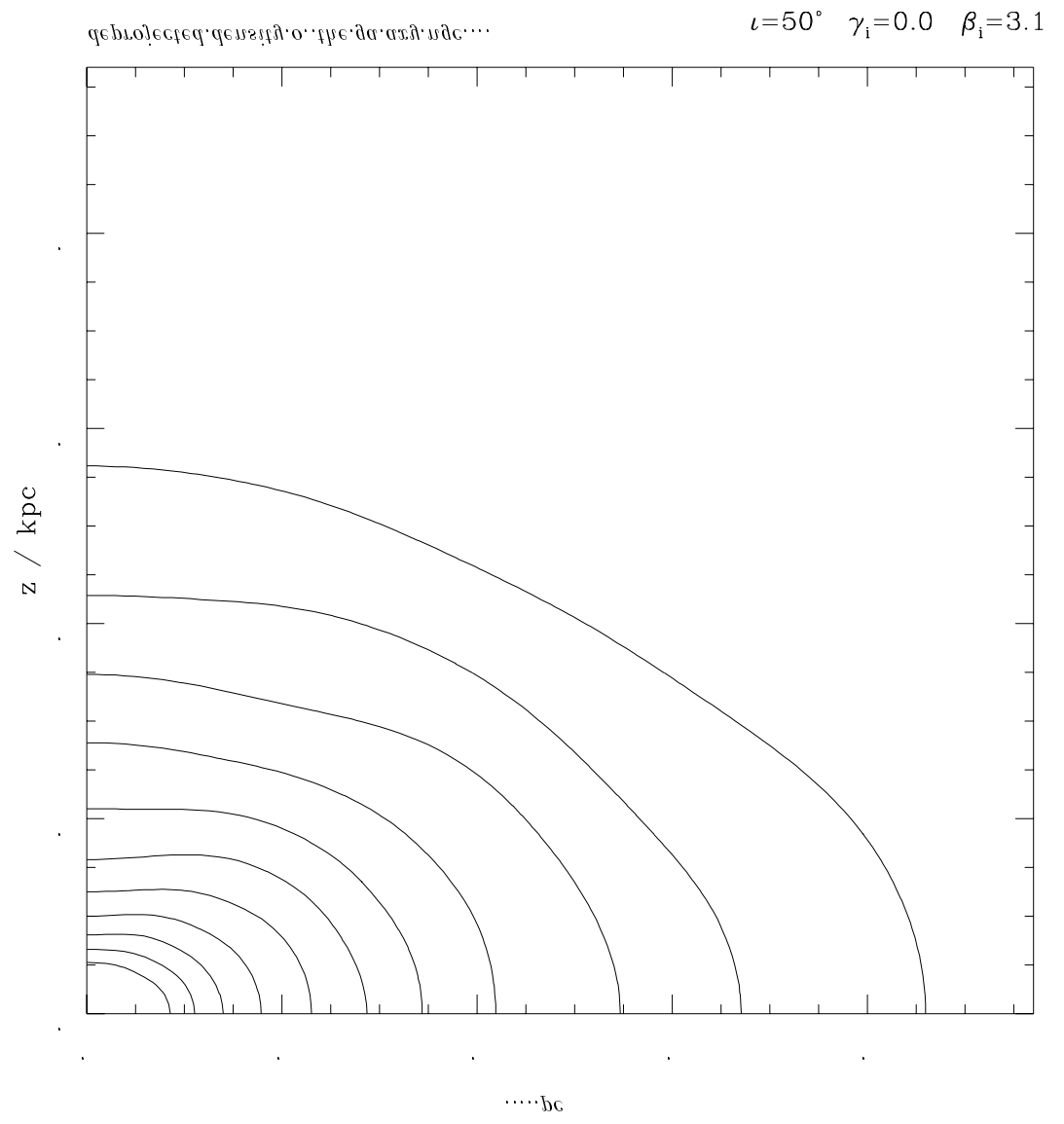

**Figure 7.** Contours of constant density in the meridional plane of a model of NGC 2300 obtained by the R–L deprojection algorithm. These are reminiscent of the sum of a konus density and an elliptical density distribution. The assumed inclination is $i = 50°$ and the rms deviations of the projected model from the data are 0.006 mag. Reproduced from Hödtke (1995).

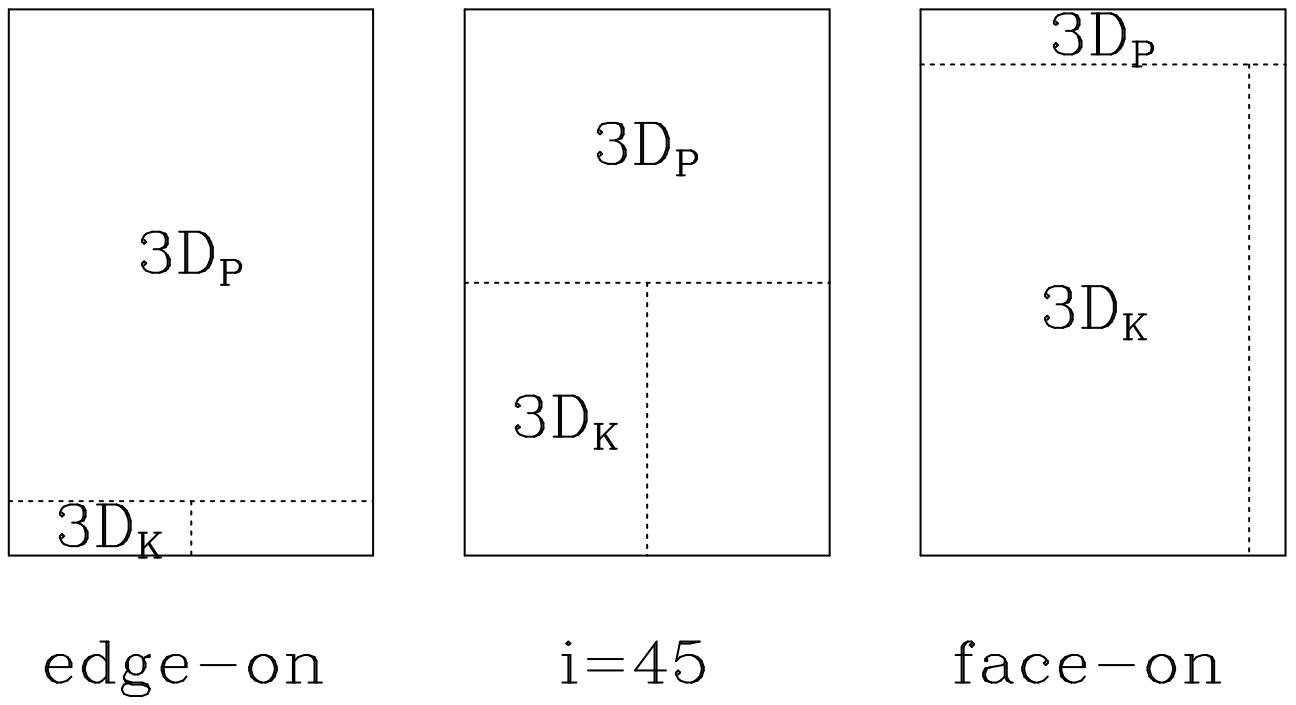

**Figure 8.** Schematic of function spaces

### 4.3 Konus densities and disky distributions

Konus densities are astrophysically important because addition of a konus density can change one's model of a given galaxy from disky to boxy or vice versa. This strongly underlines the point made by Rix & White (1990) that many elliptical galaxies might contain non-negligible disks. However, the statement of Rix & White was that these disks would be merely too faint to detect in currently available data, whereas disks contributed by konus densities could not be photometrically detected, even in principle. The amplitude of any disk that could be masked either by the addition of a konus density or by noise in the observations, decreases as the assumed inclination increases towards edge-on.

One cannot place an upper limit on the luminosity of an 'invisible' disk that can be added to a boxy galaxy without exploring the set of possible konus densities more thoroughly than we have been able to do. It seems clear that konus densities will always comprise a nested sequence of roughly conical regions of alternating positive and negative density. We have shown that some konus densities have more than the three conical regions that are characteristic of our basic family, but it seems unlikely that any have fewer. It is certain that great variety is possible in the way in which the peak value of $|\rho|$ within a conical region declines with radius in a konus density. For fixed inclination, our konus densities are characterized by two scale lengths $\alpha$ and $\beta$. The latter determines the radius beyond which the density in the equatorial plane falls as a power law, $\rho \sim R^{-3}$, while $\alpha$ sets the scale height of the disk that emerges at $z \ll R$ as $R \to \infty$. These features are certainly specific to our examples rather than characteristic of konus densities as a whole.

Galaxies are made up of stars moving collisionlessly on orbits in a given gravitational potential, and before one can feel free to add or subtract any component it is important to be assured that this can be constructed by placing stars on orbits. Fortunately, Lynden-Bell's (1962) demonstration that for any $\rho(r,\theta)$ there correspond infinitely many distribution functions $f(E, L_z)$ assures us that this is so.

In the light of the previous discussion, attempts to detect disks through their effect on kinematic observables assume greater importance. Unfortunately, uncertainty as to the role of a third integral in galaxy distribution functions makes this approach also difficult. If we knew that the distribution function were of the simple form $f(E, L_z)$, then each of the possible density distributions that are consistent with a given galaxy image, would predict different velocity profiles along the various lines of sight. Unfortunately, we have no reason to expect $f$ to be of any particular form, and the prospects for disentangling the effects of ambiguity in $\rho(r,\theta)$ and ambiguity in phase-space structure seem slim. Said differently, for each of the infinitely many densities that are consistent with a given galaxy image, there correspond infinitely many possible distribution functions $f$, and each $f$ predicts different kinematics. It seems likely that for any two different densities we can find corresponding distribution function that predict identical kinematics.

## 5 CONCLUSIONS

We have shown that for $i \neq 90°$, the deprojection of axisymmetric density distributions is non-unique in practice as well as in principle. That is, there are different, astrophysically plausible intrinsic density distributions that project to the same surface brightness.

The difference between two such densities is invisible at all sufficiently face-on inclinations. It arises from a Fourier density in the 'cone of ignorance', the region of Fourier space about which the observed image contains no information. Examples of such 'konus' densities can be constructed that are everywhere non-singular and decay rapidly with radius.

By adding a konus density to any given model of the luminosity distribution of a non-edge-on galaxy, one can change that model from disky to boxy or vice versa without, even in principle, altering the fit of the model to the photometric data. At near edge-on inclination, any added konus density will correspond to a thin disk, but the amplitude of this disk could still be significant. Thus, especially in the case of a weak disk, the disk-to-bulge ratio is photometrically undetermined.

Kinematic data will in some cases enable one to choose between luminosity distributions with identical photometric appearance. But it seems unlikely that even the most complete kinematic data will always resolve all ambiguity since the existence of a third integral for axisymmetric systems implies that an infinite number of distribution functions can generate each of the infinite number of intrinsic density distributions that are compatible with given photometric data.

The best prospect for making progress is to concentrate on studying the most edge-on systems, for which the photometric uncertainty is least. In such systems a cold disk would be unmistakable but an appropriate konus density could add to one's model a thin hot disk without affecting any observable.

We have used Palmer's deprojection algorithm to clarify the way in which the ambiguity of the deprojection, which arises from the konus densities, increases as the assumed inclination $i$ decreases from edge-on. As $i$ diminishes, the higher-order terms in the Fourier decomposition of the surface brightness affect more and more strongly the lower-order terms in the expansion of $\rho(r,\theta)$ in Legendre polynomials. This has two consequences. First, the range of surface brightness distributions that can be successfully deprojected diminishes as $i$ decreases. Second, with decreasing $i$ the deprojected density becomes more and more sensitive to noise in the data.

Although Palmer's algorithm is helpful for understanding the deprojection problem in general terms, it is much harder to program and use than the R–L algorithm. Also, it does not constrain $\rho$ to be non-negative as does the R–L algorithm.

Konus densities have infinitely many terms in their Legendre expansions. The density distribution obtained by truncating such an expansion at high order, projects to a faint surface brightness distribution. When this pattern of surface brightness is present in smoothed noisy data, a truncated konus density of large amplitude is generated when the data are deprojected. We have illustrated this phenomenon using an image of NGC 2300.

The results presented here for the axisymmetric case also have implications for general triaxial systems. For they strongly suggest that the range of triaxial densities that are consistent with given photometry is far greater than is commonly assumed. In particular, our results suggest that the three-dimensional distributions that are compatible with given photometry differ not only in the orientation and lengths of their axes, as described by Stark (1977), but also by the addition or subtraction of weaker, more local structures such as disks or dumb-bells. It would be interesting to display explicitly triaxial analogues of konus densities.

## ACKNOWLEDGMENTS

We thank R. Bender for making available the photometric data for NGC 2300 and for a stimulating discussion, and M. Hödtke for computing Fig. 7. OEG thanks the Heisenberg foundation and the Schweizerischer Nationalfonds for financial support.

## APPENDIX: BAND LIMITATION OF $\hat\rho(\boldsymbol{K})$

Axisymmetric systems with band-limited densities $\rho(\boldsymbol{x})$ evade the Fourier slice theorem because their Fourier transforms are also band-limited. That is, if $\hat\rho(\boldsymbol{k}) \equiv \int \mathrm{d}^3\boldsymbol{x}\, \exp(\mathrm{i}\boldsymbol{k}\cdot\boldsymbol{x})\rho(\boldsymbol{x})$ is the Fourier transform of $\rho$ and $(k,\theta,\phi)$ are spherical polar coordinates in $k$-space such that the axis $\theta = 0$ is aligned with the galaxy's symmetry axis, then $\hat\rho(k,\theta) = \sum_{l=0}^{L} \hat\rho_l(k)P_l(\cos\theta)$. As Palmer shows, knowledge of band-limited $\hat\rho(k,\theta)$ in the complement of Rybicki's "cone of ignorance" suffices to determine $\hat\rho(k,\theta)$ for all $\boldsymbol{k}$. In fact, $\hat\rho(\boldsymbol{k})$ is determined by its value at only a finite number of values of $\theta$ at each value of $k$.

Palmer effectively demonstrates that $\hat\rho$ is band-limited by explicit calculation. Here we show that this result follows easily from a general group-theoretic argument.

Let the functions $f_m(\boldsymbol{x})$ for $m = -l,\dots,l$ form a basis for the spin-$l$ irreducible representation of SO(3), where the action on $f_l$ of a rotation $R \in \mathrm{SO}(3)$ is

$$R : f \to f' \quad \text{with} \quad f'(\boldsymbol{x}) \equiv f(\boldsymbol{R}^T \cdot \boldsymbol{x}). \tag{A1}$$

(Here $\boldsymbol{R}$ is the rotation matrix associated with $R$.) Then by the uniqueness of the spin-$l$ representation, any $f_m$ can be written as

$$f_m(\boldsymbol{x}) = \sum_{m'=-l}^{l} f_{mm'}(r) Y_l^{m}(\theta,\phi). \tag{A2}$$

Fourier transforming equation (A1) we have

$$\begin{aligned}\hat{f}'(\boldsymbol{k}) &= \int \mathrm{d}^3\boldsymbol{x}' \exp(\mathrm{i}\boldsymbol{k}\cdot\boldsymbol{x}') f'(\boldsymbol{x}') \\ &= \int \mathrm{d}^3\boldsymbol{x}' \exp(\mathrm{i}\boldsymbol{k}\cdot\boldsymbol{x}') f(\boldsymbol{R}^T\cdot\boldsymbol{x}') \\ &= \int \mathrm{d}^3\boldsymbol{x} \exp[\mathrm{i}\boldsymbol{k}\cdot(\boldsymbol{R}\cdot\boldsymbol{x})] f(\boldsymbol{x}) \\ &= \int \mathrm{d}^3\boldsymbol{k} \exp[\mathrm{i}(\boldsymbol{R}^T\cdot\boldsymbol{k})\cdot\boldsymbol{x}] f(\boldsymbol{x}) \\ &= \hat{f}(\boldsymbol{R}^T\cdot\boldsymbol{k}).\end{aligned} \tag{A3}$$

Consequently, the Fourier transforms $\hat{f}_m$ also form a basis for a $(2l+1)$-dimensional irreducible representation of SO(3), and by the uniqueness of the spin-$l$ representation we have

$$\hat{f}_m(\boldsymbol{k}) = \sum_{m'=-l}^{l} \hat{f}_{mm'}(k) Y_l^m(\theta,\phi). \tag{A4}$$

A band-limited $\rho(\boldsymbol{x})$ is a function that can be written as a linear combination of functions that are members of spin-$l$ representations of SO(3) for $l \leq L$. By the above it immediately follows that $\hat{\rho}(\boldsymbol{k})$ has this last property and so is itself band-limited.